\documentclass[twocolumn,showkeys,aps,prb,showpacs]{revtex4-1}
\UseRawInputEncoding
\usepackage{graphicx}
\usepackage{pifont}
\usepackage[CJKbookmarks,dvipdfm,colorlinks,linkcolor=red,citecolor=blue]{hyperref}
\usepackage{graphicx}
\usepackage{amsmath, amssymb}
\usepackage{makecell}
\usepackage{multirow}
\usepackage{CJK}
\usepackage{mathrsfs}
\usepackage{bm}
\usepackage{amsmath}
\usepackage{dcolumn}
\usepackage{epstopdf}
\usepackage{dsfont}
\usepackage{amssymb}
\usepackage{tabularx}
\usepackage{array}
\usepackage{float}
\usepackage{color}
\usepackage{epstopdf}
\usepackage{mathrsfs}
\usepackage{extarrows}

\begin{document}

\title{Electric-filed  tuned  anomalous valley Hall effect in A-type hexagonal antiferromagnetic monolayer}

\author{San-Dong Guo$^{1}$, Yu-Ling Tao$^{1}$, Zi-Yang Zhuo$^{1}$, Gangqiang Zhu$^{2}$ and  Yee Sin Ang$^{3}$ }
\affiliation{$^1$School of Electronic Engineering, Xi'an University of Posts and Telecommunications, Xi'an 710121, China}
\affiliation{$^2$School of Physics and Electronic Information, Shaanxi Normal University, Xi'an 716000, Shaanxi, China}
\affiliation{$^3$Science, Mathematics and Technology (SMT), Singapore University of Technology and Design, Singapore 487372}
\begin{abstract}
The combination of antiferromagnetic (AFM) spintronics and anomalous valley Hall  effect (AVHE) is of great significance for potential
applications in valleytronics. Here, we propose a design principle  for achieving AVHE  in  A-type hexagonal AFM monolayer.
The design principle involves the introduction of layer-dependent electrostatic potential
caused by out-of-plane external electric field,  which can break the combined symmetry ($PT$ symmetry) of spatial
inversion ($P$) and time reversal ($T$), producing spin splitting.  The spin order of  spin splitting can be reversed by regulating the direction of electric field.
Based on first-principles calculations, the design principle can be verified in AFM  $\mathrm{Cr_2CH_2}$.
The layer-locked hidden Berry curvature can give rise to layer-Hall effect, including valley layer-spin Hall effect and layer-locked AVHE.
Our works provide an experimentally feasible way to realize
AVHE in AFM monolayer.

\end{abstract}
\keywords{Valley,  Antiferromagnetism, Electric field~~~~~~~~~~~~~~~~~~~~~~~~~~~~Email:sandongyuwang@163.com}

\maketitle

\section{Introduction}
 Since  the discovery and successful preparation of rich two-dimensional (2D)
materials, valley has recently received extensive attention, which  lays the foundation for
 processing information and performing logic operations  with low power consumption and high speed, as valleytronics\cite{q1,q2,q3,q4}.
Transition-metal dichalcogenide  (TMD) monolayers are typical valleytronic
materials with  a pair of
degenerate but inequivalent -K and K valleys in the reciprocal space\cite{q8-1,q8-2,q8-3,q9-1,q9-2,q9-3}.
The -K and K valleys  exhibit opposite Berry curvature and selective absorption of
chiral light. When including spin-orbit coupling (SOC),  the
valleys of -K and K points produce opposite spin splitting, characterized by spin-valley locking.
However, these nonmagnetic TMD monolayers lack spontaneous valley polarization, which hinders the wide application
of valleytronic devices.
To achieve valley splitting,  many methods have been applied, such
as external magnetic field\cite{v5,v6}, proximity effect\cite{v7,v8,v9}, light
excitation\cite{v10,v11}.

 To  realize intrinsic valley polarization,   the ferrovalley semiconductor (FVS)  has been proposed\cite{q10}, which appears in ferromagnetic (FM)
materials with broken spatial inversion symmetry. Many 2D materials have been predicted to be FVS, such as  2H-$\mathrm{VSe_2}$\cite{q10}, $\mathrm{CrSi_2X_4}$ (X=N and P)\cite{q11}, $\mathrm{VAgP_2Se_6}$\cite{q12}, $\mathrm{LaBr_2}$\cite{q13,q13-1}, $\mathrm{VSi_2P_4}$\cite{q14},  $\mathrm{NbX_2}$ (X =S and Se)\cite{q15},
$\mathrm{Nb_3I_8}$\cite{q16}, $\mathrm{TiVI_6}$\cite{q17},  FeClBr\cite{q18}.
 Realizing  valley polarization in  antiferromagnetic
(AFM) materials is more meaningful for valleytronic application, because AFM materials possess  the high storage density, robustness against external magnetic field, as well as the ultrafast writing speed\cite{v12}.
 Thus, it
is of fundamental importance and high interest to achieve valley polarization in AFM materials, accompanied by anomalous valley Hall effect (AVHE).

However, both spontaneous valley
polarization and AVHE are rarely reported in AFM monolayer. By stacking
AFM monolayer $\mathrm{MnPSe_3}$ or $\mathrm{Cr_2CH_2}$ on ferroelectric monolayer $\mathrm{Sc_2CO_2}$,  the AVHE has been achieved due to the introduction of nonuniform
potentials to break  the simultaneous time
reversal and spatial inversion\cite{v13,v14}. Spin splitting and spontaneous valley
polarization, accompanied by AVHE, have also been predicted  in AFM  Janus $\mathrm{Mn_2P_2X_3Y_3}$ (X, Y=S, Se Te; X$\neq$Y)
monolayers  by introducing an out-of-plane potential
gradient\cite{v15}.  Here, we propose a design principle  for achieving AVHE  in  A-type hexagonal AFM monolayer by the introduction of layer-dependent electrostatic potential
caused by out-of-plane external electric field. The spin order of  spin splitting can be reversed, when  the direction of electric field is flipped.  By first-principles calculations, we translate  design principle into $\mathrm{Cr_2CH_2}$ monolayer and clarify the electric-filed-tuned   valley physics.

\section{Ways to achieve AVHE}
\begin{figure}
  \includegraphics[width=8cm]{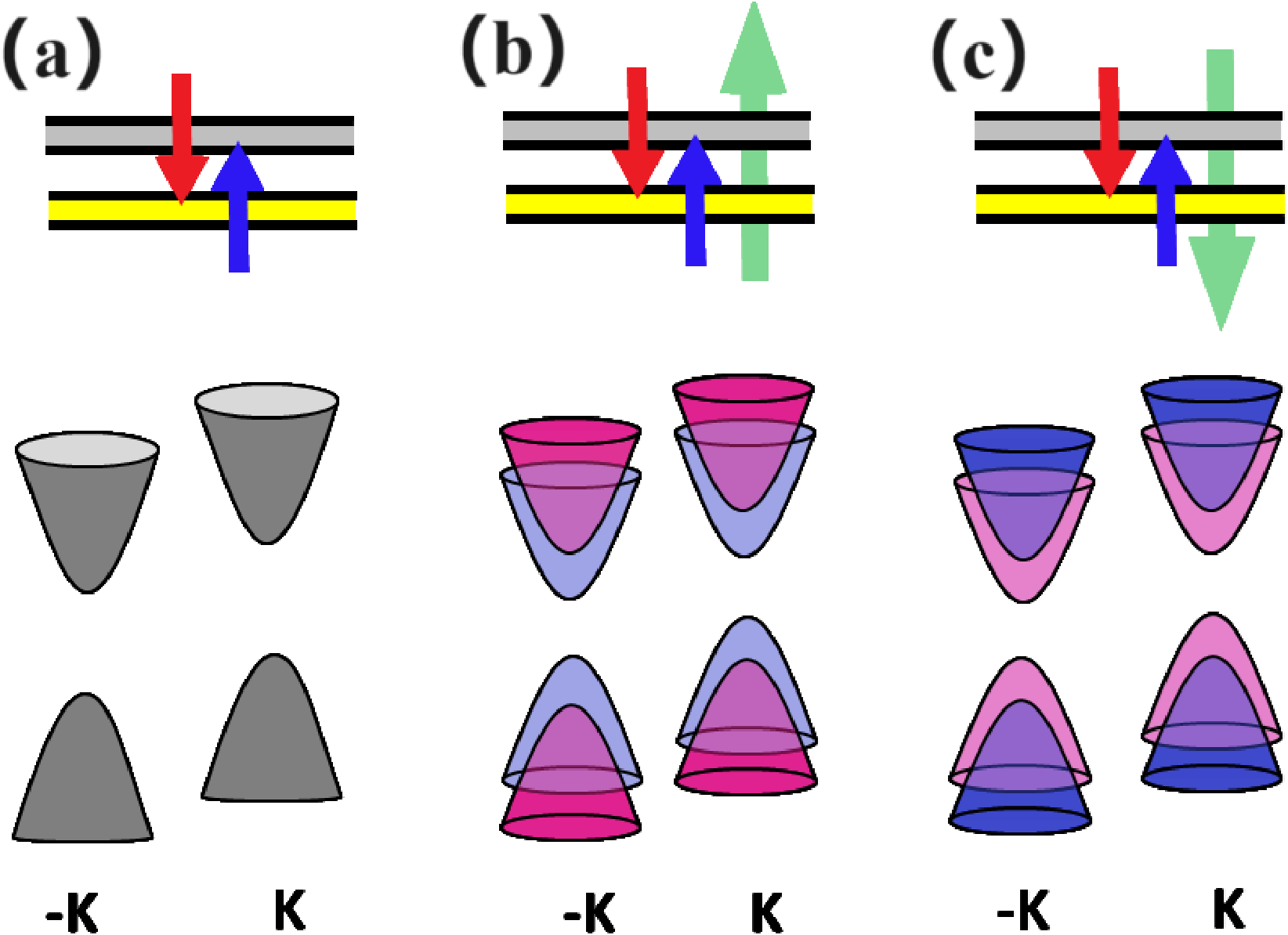}
  \caption{(Color online) (a): a hexagonal A-type AFM monolayer with spin degeneracy  but  nonequivalent  -K and K  valleys; (b): by applying an out-of-plane electric field,  the spin degeneracy is removed, and the  -K and K  valleys still are unequal; (c): when the direction of electric field is reversed, the order of spin splitting is also reversed. The spin-up and spin-down channels are depicted in blue and red. }\label{sy}
\end{figure}

The proposal for electric-filed  tuned  AVHE in
AFM monolayer is schematically illustrated in \autoref{sy}.
We consider a hexagonal centrosymmetric  monolayer with two-layer magnetic atoms with intralayer FM and interlayer AFM orderings (A-type AFM ordering), which possesses  energy extrema of conduction or
valance bands located at high symmetry -K and K points.
The lattice of our proposed system  has inversion symmetry, but  the opposite spin vectors of the two sublattices
break  spatial inversion ($P$) symmetry and time reversal ($T$)
symmetry, which produces spontaneous valley polarization (\autoref{sy} (a)). However, the the spin degeneracy of -K and K valleys is maintained  due to $PT$ symmetry, which prohibits the AVHE in AFM monolayer.

There is zero berry curvature ($\Omega(k)$) everywhere in the momentum space
 due to $PT$ symmetry.
However, layer-locked hidden Berry curvature can appear, because
each layer breaks
the $PT$ symmetry. Due to layer-spin locking,  the Berry curvatures for the spin-up and
spin-down channels are equal in magnitude and opposite in sign. The layer-locked hidden Berry curvature can give rise to layer-Hall effect.

Here, an out-of-plane electric field  is applied to break the $PT$ symmetry, which can remove spin
degeneracy of -K and K valleys.  The out-of-plane electric field can produce layer-dependent electrostatic potential $\varpropto$ $eEd$ ($e$ and $d$ denote the electron charge and the layer distance.), which induces spin splitting effect.
With breaking  $PT$ symmetry, the  spin
splitting at the -K and K valleys can be realized  (\autoref{sy} (b)), resulting in the AVHE.
 Moreover, it is expected that the spin order at both the -K and K valleys
can be reversed through manipulating the direction of out-of-plane electric field (\autoref{sy} (c)).

The H-functionalized MXene $\mathrm{Cr_2C}$ ($\mathrm{Cr_2CH_2}$)\cite{v15} can be used as   a real material to verify our proposal.
The $\mathrm{Cr_2C}$  is a half-metallic
ferromagnet, which has  been successfully synthesized in experiment\cite{v16}.
 By surface functionalization  with
H in $\mathrm{Cr_2C}$,   a ferromagnetic-antiferromagnetic transition and a
metal-insulator transition  can be induced simultaneously\cite{v15}. In addition to this,  $\mathrm{Cr_2CH_2}$ possesses A-type AFM ordering with energy extrema of
valance bands located at  -K and K points, which meets the requirements of our proposal.

\begin{figure}
  \includegraphics[width=8cm]{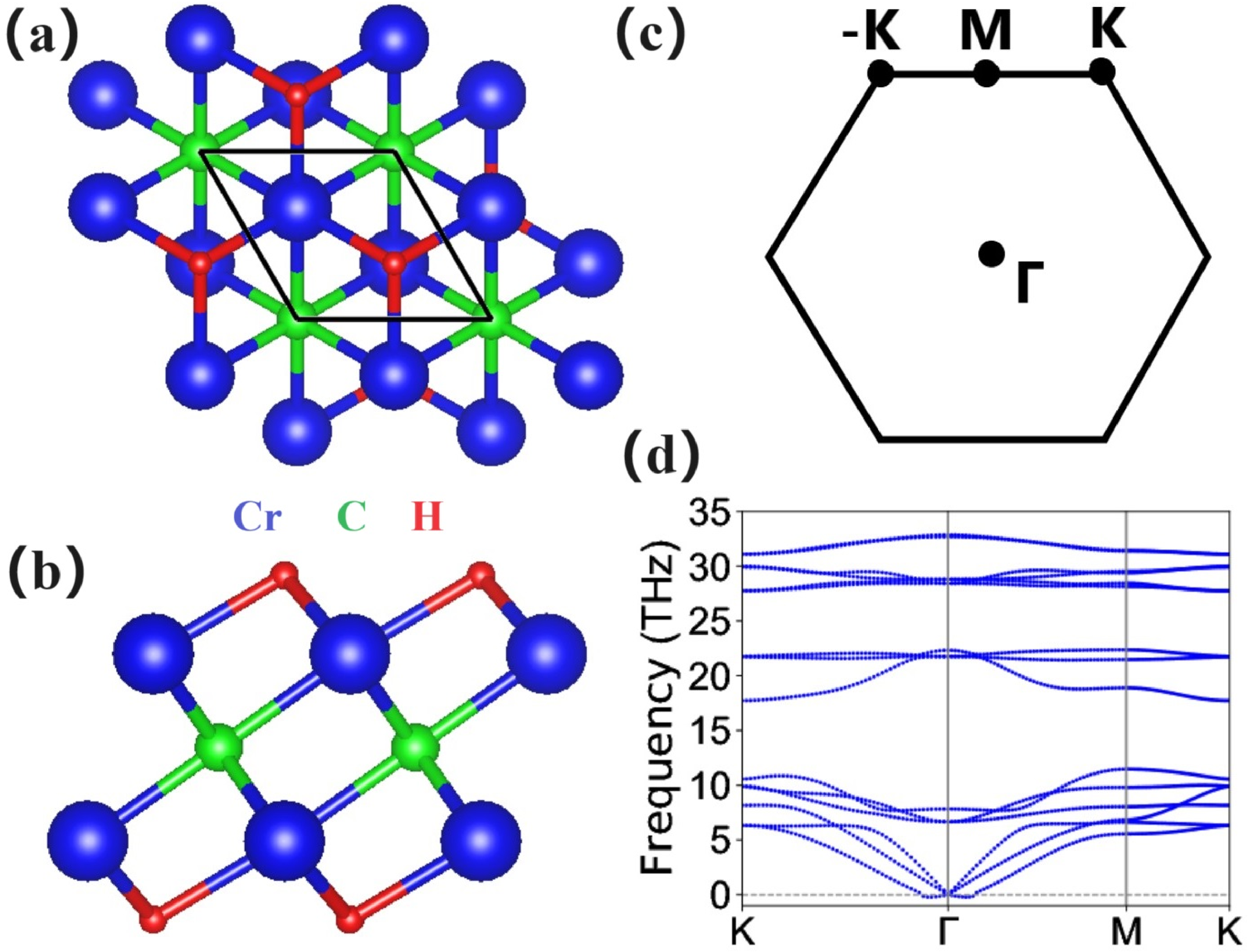}
  \caption{(Color online) For monolayer  $\mathrm{Cr_2CH_2}$,  (a) and (b): the top and side views of crystal structures; (c):  the first BZ with high symmetry points;  (d): the phonon dispersion spectrum.}\label{st}
\end{figure}

\section{Computational detail}
Within density functional theory (DFT)\cite{1}, the spin-polarized  first-principles calculations are carried out within the projector augmented-wave (PAW) method,  as implemented in Vienna ab initio Simulation Package (VASP)\cite{pv1,pv2,pv3}.  The generalized gradient
approximation  of Perdew-Burke-Ernzerhof (PBE-GGA)\cite{pbe}as the exchange-correlation functional is adopted.
The kinetic energy cutoff  of 500 eV,  total energy  convergence criterion of  $10^{-8}$ eV, and  force convergence criterion of 0.0001 $\mathrm{eV.{\AA}^{-1}}$  are set to obtain the accurate results. To account for the localized nature of Cr-3$d$ orbitals, a Hubbard correction $U_{eff}$=3.0 eV\cite{v13,v13-1}  is used by  the
rotationally invariant approach proposed by Dudarev et al. The SOC is incorporated for investigation of valley splitting and magnetic anisotropy energy (MAE).
The vacuum space of more than 20 $\mathrm{{\AA}}$ along $z$ direction is introduced to avoid interactions between neighboring slabs.
  A 21$\times$21$\times$1 Monkhorst-Pack k-point meshes are used to sample the Brillouin zone (BZ) for calculating electronic structures.
Based on  finite displacement method, the interatomic force constants (IFCs)  are calculated by employing 5$\times$5$\times$1 supercell with AFM ordering, and the phonon dispersion spectrum  is obtained by the  Phonopy code\cite{pv5}.
The Berry curvatures
are calculated directly from the calculated
wave functions  based on Fukui's
method\cite{bm},  as implemented in  the VASPBERRY code\cite{bm1,bm2}.

\begin{figure}
  \includegraphics[width=8cm]{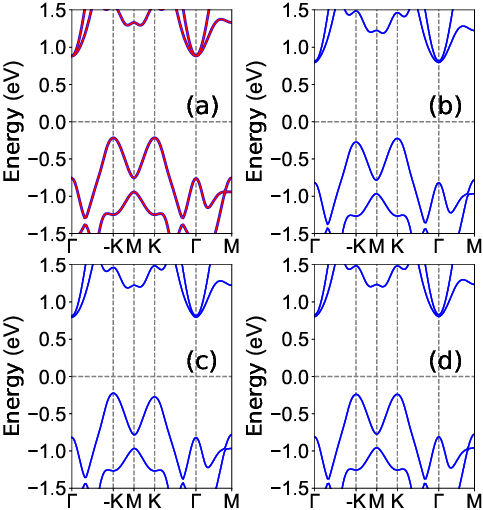}
\caption{(Color online)For  $\mathrm{Cr_2CH_2}$, the energy band structures  without SOC (a), and  with SOC (b, c, d) for magnetization direction along the positive $z$, negative $z$, and positive $x$ direction, respectively.  In (a), the blue (red) lines represent the band structure in the spin-up (spin-down) direction.}\label{band}
\end{figure}

\section{Crystal and electronic structures}
It has been proved that $\mathrm{Cr_2CH_2}$ monolayer  possesses A-type AFM ordering with  dynamical, mechanical, and thermal stabilities\cite{v15}.  The crystal structures of $\mathrm{Cr_2CH_2}$ along with the first BZ are  shown in \autoref{st} (a), (b) and (c), which crystallizes in the  $P\bar{3}m1$ space group (No.164),  hosting inversion symmetry. It consists of five atomic layers in the sequence of H-Cr-C-Cr-H,  and the magnetic Cr atoms distribute in two layers. The optimized  equilibrium lattice constants are $a$=$b$=2.99 $\mathrm{{\AA}}$ by GGA+$U$ method, which agrees well with previous result(2.95 $\mathrm{{\AA}}$\cite{v15}).  Based on its energy band structures in FIG.S1 of electronic supplementary information (ESI) or in ref.\cite{v15}, the valence band maximum  (VBM) of  $\mathrm{Cr_2CH_2}$ is at $\Gamma$ point, not -K or K point. To  clearly clarify our proposal, the biaxial strain ($a/a_0$=0.96) is applied to change VBM of  $\mathrm{Cr_2CH_2}$ from $\Gamma$ to -K or K point.  To determine magnetic ground state of  $\mathrm{Cr_2CH_2}$,  the FM and three  AFM configurations (AFM1, AFM2 and AFM3) are constructed, as  shown in FIG.S2 of ESI, and the AFM1  is called A-type AFM state. Calculated results show that the energy of AFM1 per unit cell is 281 meV, 50 meV and  399 meV  lower  than those of FM, AFM2 and AFM3 cases by GGA+$U$, confirming that the strained  $\mathrm{Cr_2CH_2}$ still possesses AFM1 ground state. The calculated phonon spectrum  of strained  $\mathrm{Cr_2CH_2}$  with  no obvious imaginary frequencies is shown  in \autoref{st} (d),  indicating its dynamic stability.

 The energy band structures of $\mathrm{Cr_2CH_2}$ by using both GGA and GGA+SOC are plotted in \autoref{band}.
According to \autoref{band} (a), no
spin splitting can be observed due to the $PT$ symmetry, and $\mathrm{Cr_2CH_2}$ is  an indirect band
gap semiconductor.
The energies of  -K and K valleys in the valence band are degenerate.  \autoref{band} (b) shows that the  valley polarization can be induced by SOC, and the valley splitting is 49  meV. The energy of K valley
is higher than one of -K valley, and  the valley polarization can  be
switched, when  the magnetization direction is reversed (\autoref{band} (c)). When the  magnetization direction of $\mathrm{Cr_2CH_2}$  is in-plane along $x$ direction (\autoref{band} (d)), no valley polarization can be observed.  Therefore, the magnetic orientation is determined by MAE, which can be obtained by $E_{MAE}=E^{||}_{SOC}-E^{\perp}_{SOC}$, where $||$ and $\perp$  mean that spins lie in
the plane and out-of-plane. The calculated MAE is 27$\mathrm{\mu eV}$/unit cell,  indicating  the out-of-plane easy magnetization axis of $\mathrm{Cr_2CH_2}$.  This confirms the realization of our proposed design principles.
 The  total magnetic moment of $\mathrm{Cr_2CH_2}$  per unit cell is strictly 0.00 $\mu_B$ with  magnetic moment of bottom/top Cr atom  being 3.09  $\mu_B$/-3.09 $\mu_B$.
When the SOC is included, the spin degeneracy is still maintained for both out-of-plane and in-plane magnetization directions.
\begin{figure}
  \includegraphics[width=8cm]{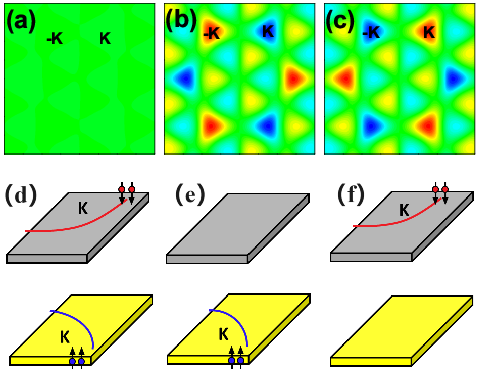}
\caption{(Color online)For  $\mathrm{Cr_2CH_2}$, the distribution of Berry curvatures of  total (a),  spin-up (b) and
spin-down (c). In the presence of a longitudinal in-plane electric field,  an appropriate hole doping for three cases in \autoref{sy} produces  valley layer-spin hall effect (d) and  layer-locked anomalous valley Hall effect (e and f). The upper and lower planes represent  the top and bottom Cr layers.}\label{berry}
\end{figure}

For  $\mathrm{Cr_2CH_2}$, the distribution of Berry curvatures of  total,  spin-up  and
spin-down  are plotted in \autoref{berry}.  It is observed that  total berry curvature  everywhere in the momentum space is zero
 due to $PT$ symmetry.
The extremes of spin-resolved Berry curvatures
locate at the -K and K valleys, which have opposite signs for the same spin channel.  For the same valley at different spin channel, the Berry curvatures are also opposite. When a longitudinal in-plane electric field is applied,
the Bloch carriers will acquire an anomalous transverse
velocity $v_{\bot}$$\sim$$E_{\parallel}\times\Omega(k)$\cite{v17}. By
shifting the Fermi level between the -K and K valleys in the valence band, the spin-up and spin-down holes from K valley will
accumulate along opposite sides of different layer under a longitudinal in-plane electric field, resulting
in the valley layer-spin Hall effect (\autoref{berry} (d)), but the AVHE is absent.

\begin{figure}
  \includegraphics[width=8cm]{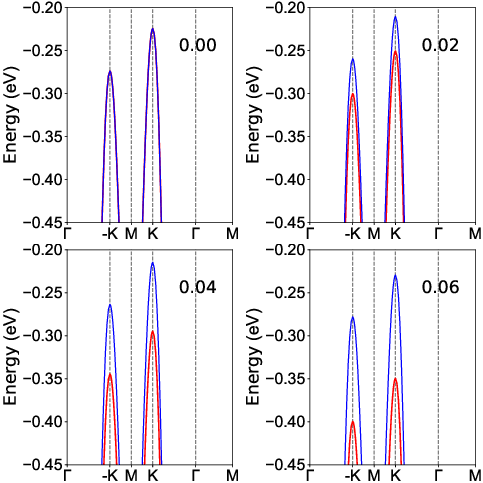}
\caption{(Color online)For  $\mathrm{Cr_2CH_2}$, the spin-resolved energy band structures near the Fermi level for the valence band with SOC at representative $E$.}\label{band-1}
\end{figure}

\section{electric field induces spin splitting}
To induce spin splitting in $\mathrm{Cr_2CH_2}$, $PT$ symmetry should be broken. Here,
an out-of-plane  electric field $+E$ (0.00-0.10 $\mathrm{V/{\AA}}$) is applied, which can break  $PT$ symmetry, and lift spin
degeneracy of valleys. The $-E$  generates exactly the same results except spin orders
since two Cr layers are related by a glide mirror $G_z$ symmetry, but have opposite magnetic moments.
Firstly,  the magnetic ground state under out-of-plane electric field  is determined by
the energy difference between  FM/AFM2/AFM3 and AFM1 configurations.
Within considered $E$ range,  based on FIG.S3 of ESI,   the AFM1 ordering is always ground state.
 The MAE vs $E$ is plotted in FIG.S4, and  the positive MAE confirms that the easy axis of $\mathrm{Cr_2CH_2}$ is out-of-plane direction within considered $E$ range. These  ensure that our design principles can be  realized in $\mathrm{Cr_2CH_2}$.

\begin{figure}
  \includegraphics[width=8cm]{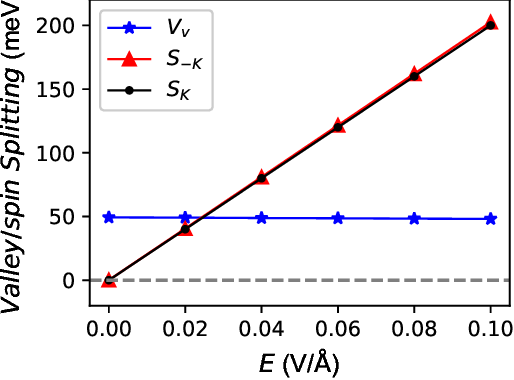}
\caption{(Color online)For $\mathrm{Cr_2CH_2}$, the valley splitting ($V_v$) and spin splitting ($S_{-K}$ and $S_{K}$ at -K and K valleys) for valence  band   as a function of $E$. }\label{ss}
\end{figure}

The energy band structures of $\mathrm{Cr_2CH_2}$ at representative $E$ by using GGA+SOC are plotted in FIG.S5 of ESI, and the  enlarged figures of spin-resolved energy band structures near the Fermi level for the valence band are shown in \autoref{band-1}.
For  valence  band,  the valley splitting  and spin splitting at -K and K valleys   as a function of $E$ are plotted in \autoref{ss}.
 The spontaneous valley polarization of about 49 meV is maintained within considered $E$ range.
When electric filed is applied, it is clearly seen that there is spin splitting,  which is due to layer-dependent electrostatic potential
caused by out-of-plane electric field. Spin splitting and electric field strength show a linear relationship. In fact, the spin splitting  can be approximately calculated by  $eEd$. Taking $E$=0.02$\mathrm{V/{\AA}}$ as a example, the $d$  of  $\mathrm{Cr_2CH_2}$ is 2.37 A, and the estimated spin splitting  is approximately 47 meV, which is close to the first-principle result 41 meV. The coexistence of  spin splitting and valley polarization is essential for the
realization of AVHE. When the electric filed changes from $+E$ to $-E$, the layer-dependent electrostatic potential is also reversed.
The sizes of spin splitting and  valley splitting  remain unchanged, but  the spin order  of spin splitting reversed (FIG.S6 of ESI), which is consistent with our proposed design principle.

For  $\mathrm{Cr_2CH_2}$ at $E$=0.02$\mathrm{V/{\AA}}$, the distribution of Berry curvatures of    spin-up  and
spin-down  are shown  in FIG.S7.  It is clearly seen that the  Berry curvatures are opposite  for the same valley at different spin channel and different valley at the same spin channel.
 By shifting the Fermi level between the -K and K valleys in the valence band, only the spin-up holes from the K valley move to the bottom boundary of the sample under an in-plane electric field (\autoref{berry} (e)), producing layer-locked AVHE.
  Conversely, by reversing  the electric field direction, the spin-down holes from the K valley will move to the top  opposite side of the sample by proper hole doping (\autoref{berry} (f)).  This accumulation of spin-polarized holes produces  a net charge/spin current, and \autoref{berry} (e) and \autoref{berry} (f) generate opposite voltage.

\section{Conclusion}
In summary,  we present a model to induce AVHE in A-type hexagonal AFM monolayer by applying electric filed.
The validity of our proposal is confirmed by an extensive study of $\mathrm{Cr_2CH_2}$ within the first-principles calculations. The spontaneous valley polarization can  occur in $\mathrm{Cr_2CH_2}$, but the spin splitting of -K and K valleys is absent.  The introduction of
an out-of-plane electric field results in  spin splitting due to layer-dependent electrostatic potential.
The layer-locked hidden Berry curvature produces layer-locked AVHE.
Our works enriches the AVHE in AFM monolayers, and provide great potential for developing
energy-efficient and ultrafast valleytronic devices.

\begin{acknowledgments}
This work is supported by Natural Science Basis Research Plan in Shaanxi Province of China  (2021JM-456). We are grateful to Shanxi Supercomputing Center of China, and the calculations were performed on TianHe-2.
\end{acknowledgments}

\end{document}